# *Exploring Indonesian Web Based Career Center*
## Discrepancy of Web Popularity and Type of Services

Renny
Department of Accounting
Gunadarma University
Depok, Indonesia
renirana@staff.gunadarma.ac.id

Reza Chandra
Department of Information Systems
Gunadarma University
Depok, Indonesia
reza_chan@staff.gunadarma.ac.id

Syamsi Ruhama
Department of Informatics Management
Gunadarma University
Depok, Indonesia
susi22@staff.gunadarma.ac.id

Mochammad Wisuda Sarjono
Department of Information Systems
Gunadarma University
Depok, Indonesia
moch_wisuda@staff.gunadarma.ac.id

*Abstract*— Utilization of the Internet in higher education focus on the learning process or the provision of academic information for students. The subject of this research is in the form of web-based management alumnus Career Center with specific sub domain. Colleges that already have a Career Center only 34 of the 264 colleges as sample. Type the service the most are information jobs, while others are still rarely available as a forum of alumni and career consultation. Ownership Career Center contributed to the popularity of college website. Providing services such as communication and consultation career impact on the popularity of the Career Center website.

## I. Introduction

The number of open unemployment in Indonesia, according to Labor statistics Indonesia amounted to 6.325%. Among the unemployed, the number of educated unemployed (unemployment Diploma and undergraduate degree) is greater than the uneducated unemployed [1]. This is understandable given the number of university graduates per year approximately 655,012 people from over 3000 universities in Indonesia [2]. Some of the things that led to high unemployment among educated is due to mismatch between the acquisition of competence education to the needs of jobs available and the quality of human resources (HR) is generated.

Added to the characteristics of students in college at the time was different in every aspect compared to ten years ago. The difference is very visible on charging administrative and faculty positions at universities including the career center position [3]. To overcome this, it is necessary to boost and strengthen the college career center to help graduates gain employment. In addition it needs to be a system that can give information to produce graduates of higher education institutions in order to ensure that college graduates have secured employment as appropriate. To facilitate career services at the college career center website needs to be created to facilitate job-seekers, especially alumni and students to view job information and online CV creation that can be seen by companies that need workers.

Five main criteria of success are: (1). A smooth transition from college to the workplace include a short waiting period of work and the search for a simple job, (2). The ratio of low unemployment, (3). The ratio of non-regular employment is low, (4). Success graduates vertically within the meaning of education investment gains or income is higher than non-graduates or graduates high ratio of working graduates. (5). The success of graduates in the sense horizontally close relationship between field of study and type of work or the high utility of knowledge gained during higher education in the work [4]. This greatly strengthens the need for research on career services center. McGrath suggests failure to provide support for the effective application of the career system may result in the placement of university graduates to be down and will put the university in a poor position / popularity decreased when competing with other universities [5].

This study aims to explore the site Career Center alumni or owned by universities in Indonesia. Research question is whether ownership and diversity is affecting the popularity Career Center and Website from college.

## II. Theoretical Background

Career center is in the structure of higher education institutions that perform functions bring together graduate students and job seekers with manual labor [2]. Career Center provides services to help students, alumni, faculty and staff fatherly career success and connect companies with students and alumni who are seeking employment.

Research on university career ce [1]nters have been conducted by several researchers [2;6;7], the results show that most students only a few use the services provided by their university career center. Meanwhile, the research shows that there has been a paradigm shift in career services that focus on delivering comprehensive services to students for undergraduate education [8]. Failure to provide support for the effective application of the career system may result in the placement of university graduates to be down and will put the university in a poor position when competing with other universities [2].

There are a number of features of distance learning that must be addressed when planning for the success of students / graduates [1]. Support services online for students is one of the features to consider. Specific suggestions for features to consider are included guidance and counseling as an essential service that should be there. Integration with other campus services are also needed as well as 24-hour access to campus services for students / alumni. Having a web page with a good performance as well as work jointly with other services available on campus is an important part of the integration [9;10].

A motivation to continue to use the site user page is based on the user's level of success. Thus research pages advocating the development process focused on meeting the needs of a specific user and the content of the charge is not the application of technology. In order to effectively implement this process, researcher shows the layout that is able to answer three questions, such as (1) Who is to serve pages? (2) What are the needs of users? And (3) What resources are available (or have available) to meet the needs of users [11].

Service career center placement and the transition from model to model of network planning. While placement and planning models can still be found, the model that dominant for career services at the moment is a network model. This model is characterized as a meeting where students, alumni, employers, faculty and staff met to handle all varieties of careers in an active association, human, print, and electronic resources career as the most efficient and effective conduct placement employment and career planning activities [12].

Services online career center also has its advantages and disadvantages. The advantage is accessible anytime and anywhere for students, easy website updates, the ability to connect and provide student referrals to relevant services organizations [13]. Utilization of web-based career services does not necessarily stop the student to ask information at the front desk [14].

The drawback is the difficulty in adapting services to the needs of individuals, finding a balance in offering flexible services to meet the exact needs of technology and human interaction can be challenging. Staff can also be a problem. With the ease of access anytime and anywhere, making the student expects no feedback at that time. Feedback is also expected to come from the smart professional career.

## III. Methodology

The sample was 264 universities in Indonesia are included in the ranking of universities based on their activity on the internet, the rank 4ICU and webometrics. Measurements on the first stage is to check whether the college has its own website for the alumni, or hereinafter called Career Center. Career center is typically an administrative unit of an organization (e.g., school, business, or agency) that employs staff who deliver a variety of career programs and services [15].

The second stage is to examine the features or services that are available in the career website, the job information, the data base of alumni, the alumni forums, consultations, tips & tricks the world of work, questionnaires alumni, external members. Career website popularity is measured using indicators reffering domain and total backlinks from www.majesticsseo.com. Contribution career website popularity is measured by the percentage of the popularity of the college website. This study also shows some examples of sites that feature career or services are complete.

Observation and measurements conducted research variables in the same week that in early January 2013. Description of the variables are presented graphically to determine the pattern of utilization of career center and the position of the Career Center for college website popularity. The influence of ownership career center websites and various career website features tested with independent sample t test and regression analysis linking with referring domains and total backlinks career with reffering web domain and total backlinks college website.

## IV. Result and Discussion

### A. Features and Services

Only 34 of 264 or 12.88 percent of colleges that have a career center with a separate sub domain address. The most widely available in the Career Center is the information that as many as 27 job vacancies from 34 web career, or 79.41%, whereas the features that are rarely available is a forum for alumni of the college was only one.

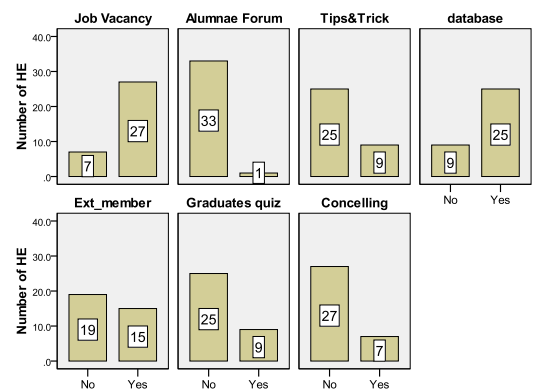

Figure 1. Type of Career Center Services

Here are 2 examples of web page display face that features a career center or most complete service.

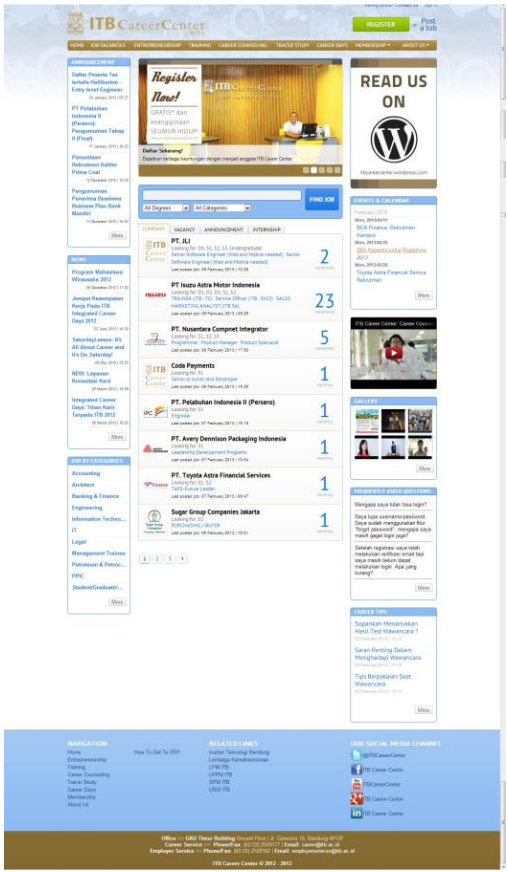

Figure 2. Career Center of Bandung Insitute of Technology (ITB)

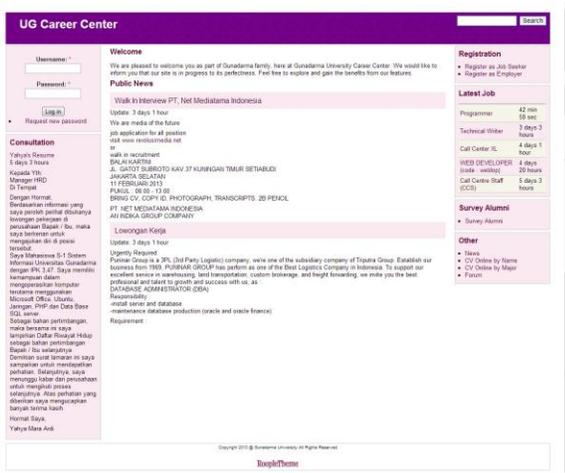

Figure 3. Career Center of Gunadarma University

The low number of colleges that have a Career Center suggests that higher education in Indonesia is relatively not use the Internet in the management graduates. The majority of colleges still use the college website for disseminating information about graduate or job. As an educational process that are sustainable, universities in Indonesia needs to improve the quality of management graduates melalalui web-based information services in the form of Carer Center. The development of career centers was one of its most tangible and lasting accomplishments [15]. Benefits Career Center was not just for graduates only, but could be from the perspective of the interests of students. Students use the one-stop resources to explore career, education, and summer job options [16].

### B. *Career Center Popularity*

Popularity Career Center reffering measured by indicators of domain and total backlinks are generally lower than the popularity of the college website. This condition can be understood as logical as the Career Center is one of the sub domains of the college website. Distribution of the number of universities by web popularity career in general follows the distribution pattern of the popularity of college website as shown in the two pictures below.

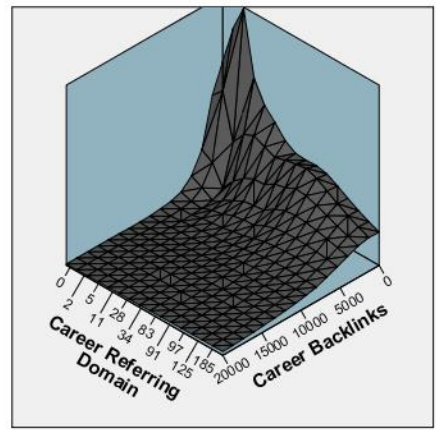

Figure 4. Sample Density of Career Center popularity

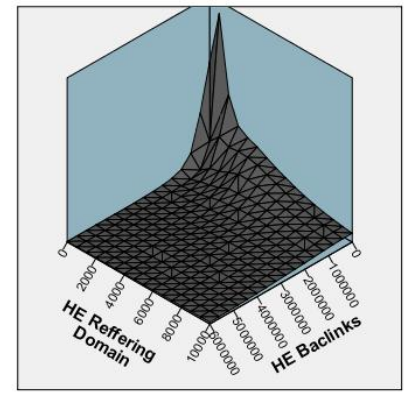

Figure 5. Sample Density of HE website popularity

Top three highest popularity career center colleges by reffering domain is Gunadarma University, University of Indonesia and Gadjah Mada University, while based on the total backlink is Gunadarma University, Binus University, and the University of Indonesia. The college was included in the top twenty on the ranking 4ICU for the January 2013 edition. Level of contribution to the popularity of the popularity of the college website college ranged from 0 to 28.18 with an average of 3.08% for reffering domain and 0 to 12.00 with an average of 1:13% for a total back links. Preview contribution to the popularity of the popularity of the Career Center website for the college domain reffering indicators can be seen in the two images below. Contribution based on total backlinks images are not presented because the level of contribution is very small.

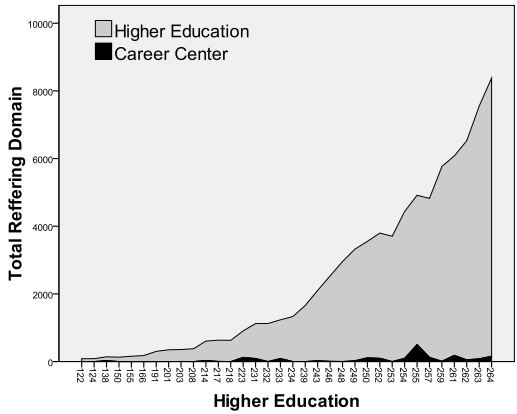

Figure 6. Contribution of Career Center referring domain

Popularity Career Center is still low suggests that the Career Center has not been widely used by web visitors, especially college graduates. The low popularity of these can technically be caused by features or services centernya Career incomplete resulting in less benefit to college graduates. Examples of information that can be used by graduates are job information, tips and tricks to enter the world of work, or consulting career. In addition, the number and high quality of content that can invite link or search results in search engines. The more tingga number or webpage content has a high chance to increase the popularity of Career center.

One of the factors that can lead to low amounts of information or documents to the Career Center, Career Center is the nature of a closed or exclusive. Enabling User ID and Password can cause the amount of information or documents is not detected by the search engines, but the policy will not reduce the number of visits or if a member of the Career Center link benefit from the service or information on the Career Center. Career Center webpage page count ranging from 1 to 47 300 pages with an average of 3914 pages, but with uneven distribution. Contributions Career Center content is still very low on the college website content with the highest contribution only 2.82%. Preview contribution amount on the Career Center web pages to the total college webpage on the website can be seen in the figure below.

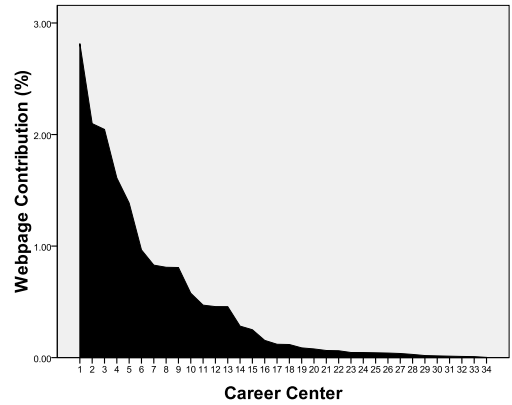

Figure 7. Webpage Career Center contribution

### C. Impact of Career Center on Popularity and Total Webpage

Measuring the impact of the Career Center consists of two parts, namely the impact of ownership on the Career Center website popularity college, as well as the impact of the type of services the Career Center to its popularity. Tests of significance using independent sample t test. The results of the first test to the table below.

TABLE I  RESULT OF FIRST TEST

|  | t-value | Significancy |
|---|---|---|
| Reffering | -10.307 | 0.000 |
| Backlink | -2.824 | 0.005 |
| Webpage | -7.004 | 0.000 |

The above test results show that there is a difference between popularity and the number of pages that have a college with a Career Center that do not. If related to the popularity of the Career Center is low, the presence of the Career Center is not the only factor causing the popularity and number of pages college. There is a tendency that the college is not popular website does not have a Career Center. This is because universities are more focused on college major domains rather than developing a special website for the Career Center.

The second test is to see the difference in popularity and the number of the Career Center webpage based on availability of services. Independent t test results can be seen in the Table II.

TABLE II    RESULT OF SECOND TEST

| Type of Service | Signifinces test (α) | | |
|---|---|---|---|
| | Referring | Backlink | Web Page |
| Job Vacancy Info | 0.168 | 0.457 | 0.304 |
| Graduates Forum | 0.000 | 0.000 | 0.000 |
| Tips&Trick | 0.267 | 0.523 | 0.816 |
| Graduates Profile | 0.091 | 0.368 | 0.396 |
| External member | 0.090 | 0.196 | 0.193 |
| Online quiz | 0.118 | 0.089 | 0.148 |
| Conselling | 0.054 | 0.057 | 0.037 |

Type of service Career Center that showed a significant difference is the forum for alumni to three indicators, as well as online conselling for total webpage. Availability of electronic forums is a great attraction for alumni to communicate or share experiences among college graduates. Information and experience from alumni is invaluable for students who also want to become a member of the Career Center. An establishment of a career centre can be a new challenge and a unique opportunity for the students to start being aware of their role and significance in the society [17].

## v. Conclusion

Universities in Indonesia have not been optimally utilize the Internet for management graduates. Some information published on the website graduates college, or not on a special site with its own sub domain. Most of the Career Center is closed is only intended for college graduates is concerned. Variety of features or types of service they provide is still not complete. The most popular type of service is a job and profile information or data base of members. Features that are rarely available is a forum for alumni and career consultation.

Ownership Career Center web-based positive impact on college website popularity. But the popularity of the Career Center itself is still relatively low when viewed from the indicators reffering domain and total backlinks. Number of content or documents on Career center is still relatively small when compared to total content of the college website. Providing a forum of alumni and career consulting the Career Center have a significant impact on the popularity of the Career Center website.

Career Development Center should be focused on the enrichment of features and content to suit the needs of alumni, or even for the community or industry. Of interest college, the Career Center can be used for data documentation alumni, graduate distribution, electronic tracer study, and form a network of alumni that can be used to share information and experiences. The development features or services in the future as proposed by Garis, Reardon, and Lenz (2012) Career advising and intake, Individual and group career counseling, Assessment and computer-assisted guidance, E-portfolio systems, Career information and networking, Career planning classes for credit, Career education outreach and programming, Web-based and onsite services, Experiential education, Career expositions, job fairs, On-campus recruiting, Job listings and résumé referral services, and Fundraising.